\documentclass{eas}
\usepackage{graphicx}

\begin{document}

\title{Disk tracing for B[e] supergiants in the Magellanic Clouds}
\author{G. Maravelias}
\address{Astronomick\'y \'ustav AV \v{C}R,
Fri\v{c}ova 298, 251\,65 Ond\v{r}ejov, Czech Republic}
\email{maravelias@asu.cas.cz}
\author{M. Kraus}
\sameaddress{1}
\secondaddress{Tartu Observatory, 61602, T\~oravere, Tartumaa, Estonia}
\author{A. Aret}
\sameaddress{2}

\begin{abstract}

B[e] supergiants  are evolved massive stars with a complex circumstellar environment. A number of important emission features probe the structure and the kinematics of the circumstellar material. In our survey of Magellanic Cloud B[e] supergiants we focus on the [OI] and [CaII] emission lines, which we identified in four more objects.

\end{abstract}

\maketitle

\section{Introduction}

The B[e] supergiants (B[e]SGs) are an important short-lived transition phase in the life of some massive stars, in which enhanced mass-loss leads to a complex circumstellar environment containing atomic, molecular and dusty regions of different temperatures and densities, presumably confined within a dense disk. Fortunately, a set of forbidden and permitted emission lines allow us to probe the properties and the kinematics of different regions in this disk. Recently, Aret et al. (\cite{Aret2012}) have identified that [CaII] $\lambda\lambda$7291, 7324 lines can be used as a  complementary observational tracer for regions forming closer to the star than the [OI] $\lambda\lambda$6300, 6363 and 5577 lines (Kraus et al. \cite{Kraus2007,Kraus2010}).

\section{Observations and Discussion}

We used FEROS (at 2.2m MPG/ESO, La Silla-Chile), during 2 observing runs (24/Nov. -- 4/Dec., 2014 \& 10 -- 18/May, 2015) to obtain high-resolution (R$\sim$48000) and wide-ranged ($\sim$3600-9200 \AA) spectra for 12 Magellanic Cloud B[e]SGs.

\begin{figure}
\centering
\includegraphics[scale=0.23]{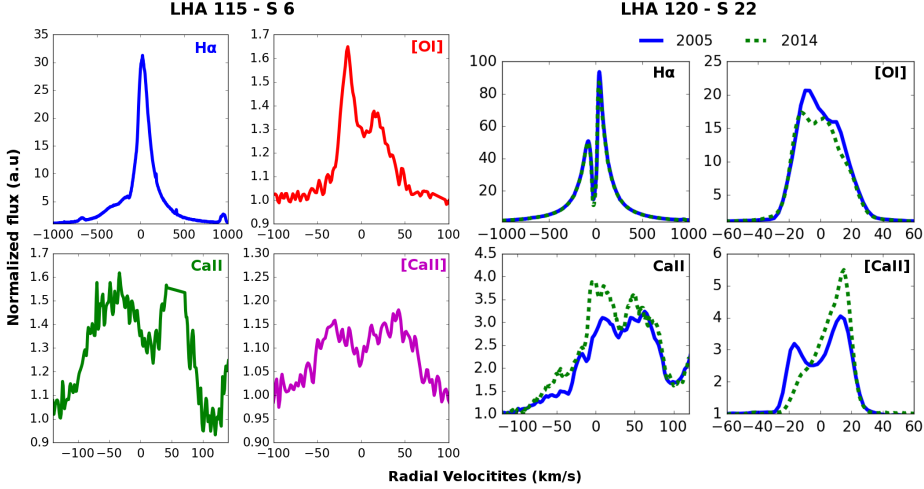}
\caption{Selected disk tracers detected in LHA115-S6 (left panel) and time variability in strategic lines of LHA120-S22 (right panel).}
\label{fig}
\end{figure}

The left panel of Fig. \ref{fig} shows selected lines of different tracers for LHA115-S6 in the SMC (new identifications in the LMC include: LHA120-S35, LHA120-S124, and ARDB 54). In all sources we identified a broad H$\alpha$ line, and emission from the CaII $\lambda\lambda$8498, 8542, 8662 triplet, a composite of wind and disk-like regions. The [OI] $\lambda\lambda$6300, 6363 and 5577 lines are present in almost all sources, being either single-peaked or asymmetric. The [CaII] $\lambda\lambda$7291, 7323 doublet is rather faint in our stars, but display double-peaked profiles, indicative of rotationally-disk structures.
 
We examined time variability of these lines, using data from 2005 (Aret et al. \cite{Aret2012}) and our own. For most sources we do not find any significant changes with respect to their spectral-line profiles and intensities. However, LHA120-S22 and LHA115-S18 show considerable changes in their line profiles. The right panel of Fig. \ref{fig} depicts the variability of LHA120-S22 indicating a highly dynamical circumstellar environment, with the characteristic transition of the [CaII] $\lambda$7291 line from double-peaked (2005) towards single-peaked (2014). 

With this work we identified the [CaII] disk tracers in 4 additional B[e]SGs in the Magellanic Clouds (almost doubling the number of studied B[e]SGs). By modeling their line  profiles we can investigate the kinematics of their formation regions. Moreover, spectra at different epochs allow us to track the evolution of the circumstellar environment of B[e]SGs and possible changes in their mass-loss behavior. 

\textbf{Acknowledgements:} 
G.M. and M.K. acknowledge financial support from GA\,\v{C}R (14-21373S) and RVO:67985815; A.A. from Estonian grants ETF8906 and IUT40-1; observations are supported by M\v{S}MT LG14013 (Tycho Brahe) project.

\end{document}